# Understanding Cost Dynamics of Serverless Computing: An Empirical Study

Muhammad Hamza[1][0000-0002-1633-9995], Muhammad Azeem Akbar[1][0000-0002-4906-6495], Rafael Capilla[2][0000-0002-6943-1285]

[1] Software Engineering Department, Lappeenranta-Lahti University of Technology, Lappeenranta, 15210, Finland
[2] Rey Juan Carlos University & Lappeenranta-Lahti University of Technology, Spain
muhammad.hamza@lut.fi, azeem.akbar@lut.fi,
rafael.capilla@urjc.es

**Abstract.** The advent of serverless computing has revolutionized the landscape of cloud computing, offering a new paradigm that enables developers to focus solely on their applications rather than managing and provisioning the underlying infrastructure. These applications involve integrating individual functions into a cohesive workflow for complex tasks. The pay-per-use model and nontransparent reporting by cloud providers make it difficult to estimate serverless costs, impeding informed business decisions. Existing research studies on serverless computing focus on performance optimization and state management, both from empirical and technical perspectives. However, the state-of-the-art shows a lack of empirical investigations on the understanding of the cost dynamics of serverless computing over traditional cloud computing. Therefore, this study delves into how organizations anticipate the costs of adopting serverless. It also aims to comprehend workload suitability and identify best practices for cost optimization of serverless applications. To this end, we conducted a qualitative (interviews) study with 15 experts from 8 companies involved in the migration and development of serverless systems. The findings revealed that, while serverless computing is highly suitable for unpredictable workloads, it may not be cost-effective for certain high-scale applications. The study also introduces a taxonomy for comparing the cost of adopting serverless versus traditional cloud.

**Keywords:** Cost Dynamics, Serverless Computing, Empirical Investigation.

## 1 Introduction

The advent of serverless computing has revolutionized the landscape of cloud computing, offering a new paradigm that enables developers to focus solely on their applications rather than managing and provisioning the underlying infrastructure [1]. Function-as-a-service (FaaS), an implementation serverless pattern, enables developers to create an application function in the cloud that automatically triggers in response to an event [1]. Companies employing the serverless model only pay for the resources consumed by the application compared to the traditional cloud, where a resource needs to be pre-reserved regardless of usage.

According to a survey conducted by Gartner Group, over 75% of organizations have either already adopted serverless computing or plan to do so within the next two years [2]. Moreover, the serverless market will substantially grow from $3 billion in 2017 to an approximate value of $22 billion by 2025 [3]. However, transitioning to a serverless computing model presents several challenges (e.g., legacy system integration, cold start, state management), and understanding the cost implications and identifying suitable workloads are crucial for effective adoption [4].

There has been significant recent research sought to address various aspects of serverless such as serverless architectural design [5], development features, technological aspects, and performance characteristics of serverless platforms [6], etc., For instance, Lin et al. [7] extensively discuss a serverless architecture, proposed a formal construct for defining serverless application workflows, and introduced the Probability Refined Critical Path Greedy algorithm (PRCP) to optimize both performance and cost. Also, Wen et al. [8] conducted a systematic literature review and highlighted the benefits of serverless computing, its performance optimization, commonly used platforms, research trends, and promising opportunities in the field. However, to the best of our knowledge, no empirical study extensively investigated the systems transitioned to serverless computing or greenfield development. This includes aspects such as predicting serverless cost, serverless workload applicability, and cost optimization. Furthermore, there is a lack of taxonomy to compare the cost of adopting serverless and traditional cloud computing.

Therefore, this study investigates companies' decision-making process to determine the cost-effectiveness of adopting serverless computing. It also evaluates the suitability of various workloads for serverless computing. Additionally, the research identifies factors that contribute to high costs in serverless applications and explores the practices to optimize them. To this end, we analyzed eight systems that have successfully transitioned to serverless computing by conducting 15 interviews with industry professionals. In addition to our empirical analysis, we developed a taxonomy for comparing the cost of adopting serverless and traditional cloud computing.

Following, we presented three research questions that guided our study:
**RQ1:** How do companies estimate the cost of adopting serverless computing?
**RQ2:** Which specific types of workloads are best suited for serverless computing?
**RQ3:** What factors may increase the cost, and how can they be optimized?

The paper is structured as follows: Section 2 delves into related work, Section 3 outlines the research method, Section 4 discusses the results, Section 5 introduces the taxonomy on cost components, and Section 6 concludes the study.

## 2    Related Work

The existing studies have discussed different aspects of serverless computing, including architectural design, performance improvement, technological aspects, testing and debugging [9] [10], and empirical investigations [11], [12], [13].

Wen et al. [11] analyzed 619 discussions from the stack overflow repository. Their study uncovered the challenges (e.g., function configuration, package integration, function invocation) that developers face when developing a serverless application. Similarly, Eskandani and Salvaneschi [12] provided insight into the FaaS ecosystem by analyzing the 2k real-world open-source applications developed using a serverless platform. The study collected open-source applications from GitHub and explores aspects like the growth rate of serverless architecture, architectural design, and common use cases. A similar study conducted by Esimann et al. [13] analyzed 16 characteristics that described why and when successful adopters are using serverless applications, and how they are building them by analyzing GitHub serverless projects [12].

Additionally, Adam et al. [14] propose guidelines for migrating to FaaS, aiming to optimize serverless functions to reduce memory consumption and running costs by conducting local experiments with their application. Another study conducted by Tarek et al. [15] developed an algorithm to optimize the cost of serverless applications through function fusion and placement. Similarly, Anil et al. [16] evaluated the AWS (Amazon Web Services) step function orchestrator concerning its performance and cost by conducting a series of experiments. Adzic and Chatley [17] conducted two industrial case studies from early adopters, demonstrating how transitioning an application to the Lambda deployment architecture reduced hosting costs. Their study did not present the cost optimization practices for companies.

Our study differs from the previous ones as we empirically investigate how organizations anticipate the cost implications of serverless computing. It also evaluates the suitability of various workloads for serverless computing. Additionally, the study identifies factors that contribute to high costs in serverless applications and explores the practices to optimize them. The existing studies did not cover these aspects of serverless computing.

## 3  Research Methodology

We employed a qualitative research method, specifically semi-structured interviews [18], to fulfill the objective of this study. Qualitative approaches aim to understand real-world situations, deal directly with complex issues, and are useful in answering "how" questions in the study [18]. The interviews were undertaken with 15 industrial participants who have experience in migrating legacy systems to serverless architectures or in developing serverless systems from scratch.

### 3.1  Data collection

*Interview Instruments.* The semi-structured interview guide was developed based on the research questions following the guidelines of Robinson [19]. The interview guide covers demographic information, strategies followed by companies to understand the cost dynamics of serverless, serverless workload applicability, and strategies for optimizing application cost. The first and second authors were involved in developing the interview questions. The interview guide can be found at[1].

*Participants Recruitment.* The first two authors attended seven technology innovation industrial meetups where companies participated to share their success stories. Both authors randomly contacted industrial practitioners and asked them whether they employed serverless computing in their industry. In addition, the second author contacted the targeted population by leveraging social media platforms (e.g., LinkedIn, ResearchGate). A total of 38 participants were contacted, of which 15 were selected for the interview. We adopted a defined set of acceptance criteria for selecting our interviewees and case organizations. Mainly, our participants are (a) professional software engineers (b) who have participated in a serverless migration project within their professional scope or developed greenfield serverless application.

We finally shared the interview script with the practitioners beforehand to familiarize ourselves with the study. We interviewed 15 professionals from 4 countries (Finland, Netherlands, UAE, Pakistan) working at medium and large companies in different business domains. The first author conducted all the interviews online using Zoom and Microsoft Teams platforms. The interviews lasted for ~40 to ~55 minutes on average. The recorded interviews were transcribed for further analysis.

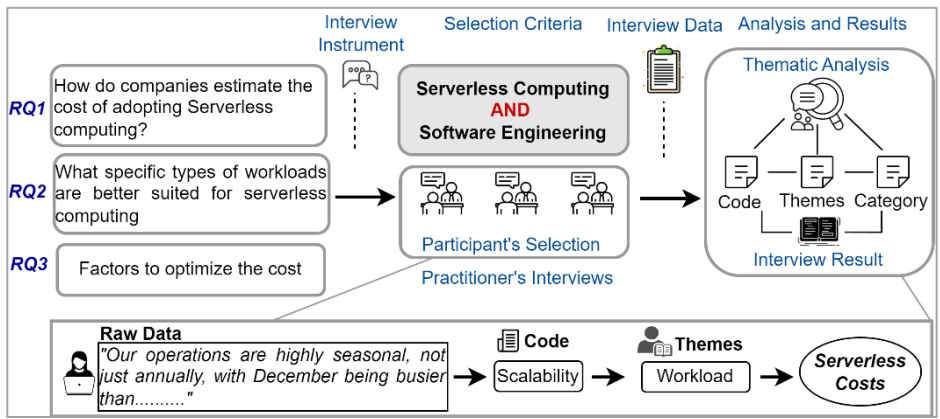

**Fig. 1.** Research Methodology

### 3.2  Data Analysis

This study used a thematic analysis approach to identify, analyze, and report the findings [20]. The thematic analysis enabled us to identify decision-making practices, workload applicability, and cost optimization practices, which were subsequently mapped into themes. We utilized NVivo[2] qualitative data analysis tool to identify and categorize the codes into themes. Initially, we meticulously read the interview transcriptions and made observational notes without establishing codes. After familiarization, we began coding the transcriptions, scrutinizing, and categorizing the resultant codes under the main themes. The main themes were decision, workload applicability, and cost optimization. The coding part was revisited repeatedly, and statements with similar meanings, but different phrasing were connected.

---

[1] https://tinyurl.com/2kdraumf
[2] https://support.qsrinternational.com/s/

**Table 1.** Company's demographics.

| Company | ID | Domain | Employees |
|---|---|---|---|
| Co.1 | S1 | Logistics services | 37365 |
| Co.2 | S2 | E-commerce | 15000 |
| Co.3 | S3 | Web-applications | 14500 |
| Co.4 | S4 | E-commerce | 9500 |
| Co.5 | S5 | E-commerce | 700 |
| Co.6 | S6 | AI & Security Services | 536 |
| Co.7 | S7 | Smart mobility and security | 20 |
| Co.8 | S8 | E-commerce | 3500 |

## 4 Results and Discussion

We conducted a comprehensive thematic analysis to obtain our results. Codes were extracted from interview transcripts and subsequently mapped into themes. These codes are denoted as C1, C2, C3, etc., while the corresponding themes are labeled T1, T2, and T3. Figure 2 provides a detailed representation of all identified codes and themes.

### 4.1 T1: Estimating Serverless Cost (RQ1)

In this section we present the practices practitioners employ to assess the cost of adopting serverless computing. Companies conduct a thorough cost analysis comparing the current infrastructure costs with the projected costs of serverless architecture. The following are the strategies reported by interviewed participants to predict the cost of serverless.

*C1: Understanding Systems Nature.* Serverless charges based on the pay-per-use model as compared to the traditional cloud. Therefore, understanding the nature and workload of the system is crucial before adopting a serverless model. The interviews revealed that serverless is the best fit for a system that receives a highly unpredictable workload. Many of the systems investigated follow an event-driven style. For instance, participant P1 stated, *"Our operations are highly seasonal, not just annually, with December being busier than June, but [...]. Given this variability, a serverless, event-driven architecture makes sense. It scales with the events, and we only pay for the events we use, reducing costs during off-peak times"*. In such scenarios, companies are compelled to over-provision each service, resulting in substantial resource wastage due to unused CPU utilization. Therefore, our interviewed participant assisted in assessing the workload of the system and monitoring the resource utilization of servers to decide to adopt serverless P1 further stated, *"It's quite costly, and it genuinely pains me to witness an AWS account operating hundreds of EC2 instances, each running at less than 5% CPU utilization"*.

*C2: Focusing on unit economics*. Unit economics can guide the decision to adopt serverless models by comparing the cost per unit of request between current and serverless architectures. In this case, 8 out of 15 participants agreed that doing the unit analysis can help make informed decisions for adopting serverless in terms of cost-effectiveness. If serverless offers a lower cost per unit, it may be a cost-effective choice P3 stated, *"I've realized the importance of understanding the unit economics of the systems we build. By identifying the cost per unit of value - for instance, the cost per scan in a security website scanning system - we can better manage resources and demonstrate our true profitability. This approach is particularly beneficial in serverless architectures".* Another participant P8 stated that *"Based on my calculations, handling 100 million requests via API Gateway and Lambda is cost-effective and more scalable compared to traditional clusters."*

*C3: Testing costly components with serverless*. participants identified the most expensive components in a large monolithic system and employed domain-driven design to extract these components. They migrated these isolated components to a serverless architecture to assess whether this transition is cost-effective. For instance, P9 stated, *"We advocate for serverless rightsizing. We start by identifying the most expensive components in a legacy system and strategically migrating them to a serverless architecture. An automated cost-benefit analysis accompanies this process, providing solid justification for the transition. In our experience with serverless, we've seen the potential for substantial returns, even up to a 100-fold return on investment".* Therefore, testing the costly component with serverless and gradually migrating is the best practice reported by the participants to be cost-effective.

*C4: Enabling a cost-conscious team.* Empowering a cost-conscious team is a crucial step in evaluating the cost implications of adopting serverless architecture and making an informed decision about the serverless in terms of cost-effectiveness. As stated by P13: *"So you know, you need someone who understands both the finance side of things, as well as the technical side of things to really sort of kind of appreciate some of the total cost of ownership applications that serverless has".*

*C5: Serverless first mindset*. organizations developing greenfield projects must go with a serverless first mindset P15 stated: *"I think if you're a startup and you're building on AWS, it just doesn't make sense for you to do anything than serverless [...] You know, the cost of containers is so much more operations work, and probably must hire some specialists, just to look after your container environment".* However, applications having high throughput could not be cost-effective in serverless computing as stated by P14 *"The funny thing is that a lot of the enterprises, they don't really have that high throughput applications where you will be significantly more expensive to run on serverless compared to containers".* However, to effectively understand the cost-effectiveness of serverless computing, it's crucial to deeply understand the nature of the system, emphasizing on unit economics, assessing the costly components of legacy application, and testing with serverless, and cultivating a team that is acutely cost-aware.

### 4.2 Interview Cases Description (RQ2)

This section delves into the case studies of systems that have either migrated to a serverless architecture or were developed greenfield serverless systems. We investigated eight systems by interviewing 15 participants, which we refer to as 'S1-S8,' from companies labeled as 'Co.1-Co.8' (where 'Co' stands for 'Company' and 'S' stands for 'System'). The details of participating companies (Co.1-Co.8) of different sizes and domains are shown in Table 1 and Table 2. We presented a short introduction to each system naming them S1-S8 from companies Co.1-Co.8. Furthermore, we understand the type of traffic the systems were receiving (e.g., unpredictable, or spiky traffic, predictable traffic). We derived three codes (C6: unpredictable or spiky workload, C7: workload having less than 1000req/s, C8: predictable workload) by analyzing the eight systems and mapped into themes T2: workload applicability presented in Figure 2.

Table 2. Participant's demographic

| Participants | Participant's Role | Professional Experience | Serverless Experience |
|---|---|---|---|
| P1 | Architect | 18 | 5 |
| P2 | Architect | 16 | 5 |
| P3 | Architect | 9 | 3 |
| P4 | Architect | 13 | 5 |
| P5 | Developer | 8 | 3 |
| P6 | Architect | 15 | 4 |
| P7 | Lead Engineer | 5 | 2 |
| P8 | Software Engineer | 5 | 2 |
| P9 | Team Manager | 5 | 2 |
| P10 | Software Engineer | 6 | 2 |
| P11 | Architect | 18 | 5 |
| P12 | Architect | 16 | 5 |
| P13 | Architect | 9 | 4 |
| P14 | Architect | 13 | 5 |
| P15 | Developer | 8 | 4 |

*Co.1-S1 Logistic Management System.* Co.1 is a large-scale enterprise offering logistics services, including domestic and international mail and parcel delivery and e-commerce solutions. The system was facing seasonal traffic, causing the organization to handle the underlying operational overhead. P1 stated that: *"Our operations are highly seasonal, not just annually, with December being busier than June, but also weekly and daily. For instance, Tuesdays are busier than Mondays, and there's a surge of traffic around 4:00 p.m. and 5:00 p.m. A serverless architecture scales with events and cuts costs during off-peak times, [...]"*. This company first evaluated the system's nature and then conducted a proof-of-concept (POC). Additionally, they identified the expensive components in a traditional cloud setting and tested them with a serverless approach. The company was able to cut costs by 80 percent and reduce delivery times from months to minutes for its e-commerce API services migrating to serverless P1 stated, *"The business case became evident when we realized that by transitioning from a fixed instance and discarding our old data-management software, we could reduce our data-management platform costs by at least 80%"*.

***Co.2-S2 E-Commerce.*** The company simplifies daily life for thousands of satisfied customers by offering a wide range of products for everyday needs and special occasions. They offer delivery at a time that suits the customer, often on the same day. According to P2: *"So we have very low traffic at night, steady traffic during the day, small spike at lunch, goes up in the evening, and then it dies off at midnight."* So, the system faced seasonal traffic in peak times and was facing challenges managing servers. They extracted components from the legacy application and tested with serverless. They did the unit calculation of the received traffic and decided serverless could reduce the cost and improve the scalability. The migration reduces significant costs and operation overhead.

***Co.3-S3 Digital Product Development.*** The company offers a variety of digital services designed to help businesses thrive in a digital-centric landscape using their web-based platform. The company has predictable traffic, handling millions of requests per month and wanted to reduce the operational overheads. They leveraged the serverless and reduced the cost from 1 thousand dollars to five hundred as stated by P7: *"By migrating from EC2 to serverless, we drastically reduced our costs while still providing the same services"*.

***Co.4-S4 Pitch Decker.*** This company helps startups with various aspects, such as pitching to investors and getting up and running. Initially, they used AWS EC2 instances for hosting but encountered scalability and maintenance issues. P4 stated: *"We struggled with determining when to scale up or down as our app, not being time-sensitive or event-driven, didn't present predictable traffic spikes [...]"*. They were spending a lot more time managing the underlying infrastructure rather than focusing on the business logic. Therefore, migrating to serverless reduced the operational overhead as the company does not want to hire a DevOps team.

***Co.5-S5 E-commerce.*** The company specializes in providing custom apparel and accessories to its customers using its design tools. The company was facing the high cost of managing the servers and scalability issues as they received unpredictable workloads during the seasonal time stated by P5: *"We had to move that to a sort of more performance, more scalable system, where we didn't have to sort of keep scaling up these EC2 instances"*. They moved a key part of their design architecture from an app to a Node-based Lambda. This transition resulted in 90% cost savings and improved performance and scalability. *"We got like immediate cost savings as well as sort of a capability expansion"*.

***Co.6-S6 AI Virtual Assistant.*** The company provides financial services with artificial intelligence and machine learning (AI/ML) solutions. The system can read, comprehend, and draw conclusions based on context to mimic cognitive thinking and build expertise over time. Their previous infrastructure (EC2) was becoming increasingly expensive, with their monthly cloud bill rising. The system consistently manages a steady and predictable volume of traffic. However, migrating to serverless reduced the cost significantly, as stated by P12: *"After assessing the serverless pay-per-use model, we opted to implement it, resulting in an impressive cost reduction of approximately 87%"*.

***Co.7-S7 Smart Mobility System.*** The startup company developed a smart mobility data generation system. This system involves collecting data from mobile phones and sending it to the startup's backend infrastructure. The startup wants to develop a system where they reduce the cost of the system and does not manage underlying infrastructure, as stated by P10: *"The need for scalability and flexibility in their operations was paramount. We want to get rid of like the time we spent on managing servers"*. The company evaluated that the nature of its system is event-driven and will grow exponentially, so it decided to go with a serverless first mindset.

**Co.8–S8** *E-Commerce.* The company provides e-commerce services mainly for ordering food and grocery items. Initially, the company had a big monolithic system and faced issues such as scalability during peak seasons as their traffic was unpredictable, faster time to market, and high operation overhead (e.g., managing EC2 instances). These issues led to increased costs. The P11 stated that *"We wanted to create something we could own and rapidly iterate on. However, I was concerned about scaling and didn't want to deal with potential EC2 server crashes or backend container issues"*. However, migration to serverless improved the scalability and reduced operational overhead and overall cost significantly.

Most of the interview systems (5) and participants (11) reported that migrating the unpredictable or spiky workload to serverless would significantly reduce the cost. However, three systems had a predictable workload and stated that they reduced the cost of going serverless P9: *"While running containers might seem cheaper initially, the hidden costs of expertise, maintenance, and scalability can quickly add up. Serverless, despite a potentially higher bill, can save costs by eliminating the need for specialized skills and infrastructure management"*. So, there is a tradeoff going serverless. Six out of 15 participants agreed that there are no universal solutions, only tradeoffs, and the choice between serverless and containers depends on the specific context and requirements. While serverless theoretically offers infinite scalability, it has a burst concurrency limit stated by P13 *"you know at high scale (1000+ req/s), services like API Gateway and Lambda can be more expensive than running containers on ECS. Lambda may also not be suitable for long-running tasks that take more than 15 minutes or applications with strict latency requirements"*, making it unsuitable for certain stabilized high-scale applications.

### 4.3 Cost Optimization Practices (RQ3)

This section highlights the primary factors increasing the costs in serverless architecture and outlines some solutions to optimize these costs from the practitioner's perspective.

*C9: Recursive Function Calling.* Refers to the situation where a serverless function triggers itself, directly or indirectly, causing a loop of invocations. This recursive triggering can result in many function invocations, increasing the overall computation time and potentially leading to unexpectedly high costs. Practitioner P 6 stated: *"During our work with a customer's system migration, an unexpected cost spike occurred due to code calling the KMS API millions of times, which they were unaware of until we generated an alert"*. However, practitioners employ different practices, including error handling and retry policies, use of idempotency keys, circuit breaker pattern, rate limiting, and recursive loop detection to handle the recursive function calling.

*C10: Unused functions.* functions deployed but not invoked or used over a significant period occupy resources and may incur costs even if they are not actively serving requests. According to P8: *"We periodically review and delete unused Lambda functions and associated resources (e.g., API Gateway, DynamoDB tables, S3 buckets) to minimize unnecessary costs"*.

*C11: Unintended logging.* refers to excessive log data generation due to debug-level logging, verbose logging, or configuration mistakes. This not only incurs unnecessary costs for data storage and transfer in services but also complicates the process of extracting useful information from the logs. *"We experienced excessive data collection in monitoring solutions like Datadog that lead to significant costs, especially as usage scales from development to production [...] "*.

*C12: Inefficient data access patterns*. This leads to a situation where developers might store a relatively small amount of data external database, but they're accessing or retrieving that data frequently. If the data is being retrieved millions of times a day, even if it's a small amount, the costs for these API requests can add up quickly and become significant. P11 stated: *"Inefficient access patterns in S3, such as frequent API calls to retrieve small amounts of data, can significantly increase costs, even if the stored data volume is low"*. Our interviewed practitioners mitigate this problem by considering data access patterns and optimizing them to minimize the number of API requests. This might involve using caching, batch retrieval of data, or redesigning their application to reduce the frequency of data retrieval.

*C13: Denial of wallet attack*. In this attack, an attacker intentionally triggers many function executions in a serverless application to inflate the application's operational costs. According to P9: *"We're aware of the risk of Denial-of-Wallet attacks in serverless architectures. Rapid scaling can lead to significant costs, so we ensure to have alerts and alarms in place to prevent unexpected expenses"*.

Our interview revealed the practices that need to be adopted to optimize the cost of serverless applications.

*C14: Function Right Sizing.* Involves matching the allocated resources to the actual usage of your functions. Over-provisioning can lead to unnecessary costs, while under-provisioning can hurt performance, as stated by *"We've learned that finding the 'right sizing' for Lambda functions is crucial - balancing performance and cost by continuously fine-tuning settings like memory allocation"*.

*C15: Provisioned Concurrency.* Keeps functions initialized and ready to respond instantly for reserved instances. However, mismatching the reserved instances can lead to high cost. "We prioritize optimizing cost and performance in operations […] understanding concurrency patterns and behavior is essential for effective implementation".

*C16: Observing System Metrics.* System metrics can provide insights into the application's performance and resource usage. This information can guide optimization efforts and help identify potential cost savings. According to P9*: "You just keep an eye on things, make sure that you haven't missed any alerts or stuff like that, which is great when you've got talking about the system of time and it being an operational thing for cost data because there's such a big delay"*.

*C17: Direct Integration.* Involves connecting services directly instead of using intermediary services. This can reduce latency, improve performance, and lower costs *"I *have personally witnessed the advantages of directly integrating serverless services, which can effectively decrease Lambda costs"*.

*C18: Avoiding idle time.* Refers to the period when resources are allocated but not actively used. In a serverless architecture, you're billed for the computing time you consume, so reducing idle time can significantly cut costs. *"We know it's vital to avoid idle wait time in Lambda functions; using Lambda as an orchestrator for long gaps incurs unnecessary costs, so we optimize by focusing on active processing tasks"*.

Apart from these, practitioners also highlighted that optimizing the code of the function, enabling billing alerts, giving developers billing access, and evaluating third-party tooling can significantly improve the optimizations and cost of the serverless application.

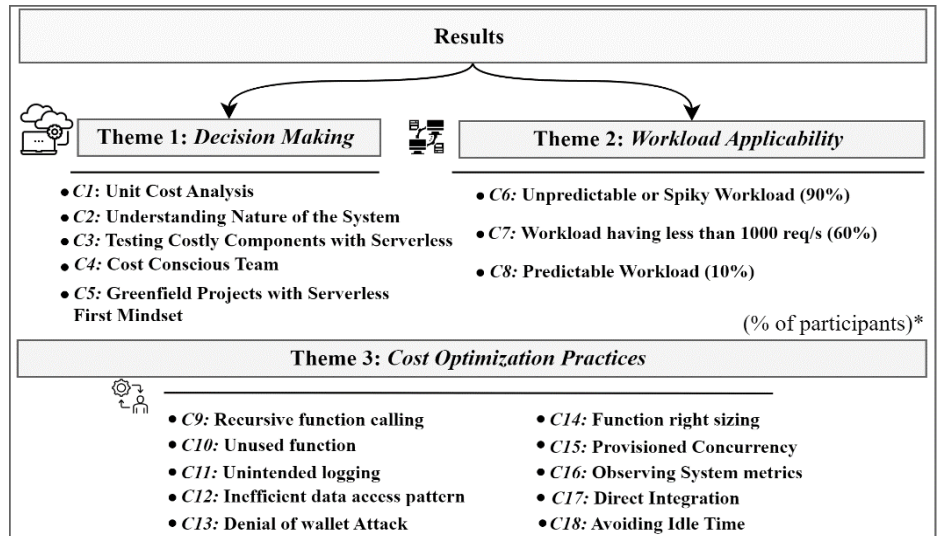

**Fig. 2.** Results from thematic analysis

## 5   Taxonomy of Factors Comparing the Cost of Ownership

In this section, we presented a taxonomy of factors comparing the cost of ownership between serverless and traditional cloud computing. The model is mainly divided into three components (i.e., infrastructure, development, and maintenance). We explained these components in detail and compared them with serverless and traditional cloud. This comparative analysis aims to provide organizations with insights to make informed decisions, comparing their cost of ownership in either computing model.

*Infrastructure cost.* Incurred when utilizing a cloud service provider for hosting an application workload. The infrastructure cost comprises the computing, storage, and network services the host application consumes. On the traditional cloud, the computing cost is calculated based on the reserved instances for a specific period, whereas in serverless computing, the cost is calculated by actual execution time, achieving the 100% utilization of the resources. Our empirical analysis showed that systems on EC2 instances or servers were not fully utilizing their computational resources leading to waste of resources and operational overhead. Furthermore, utilizing services such as load balancing, fault tolerance, and security cost extra charges on the traditional clou, whereas serverless architecturally provides these services. Organizations further need to evaluate the cost of database (e.g., compare the cost of querying NoSQL, such as MongoDB and DynamoDB). Therefore, organizations need to compare the computing, storage, and network cost of serverless and traditional cloud to make an informed decision.

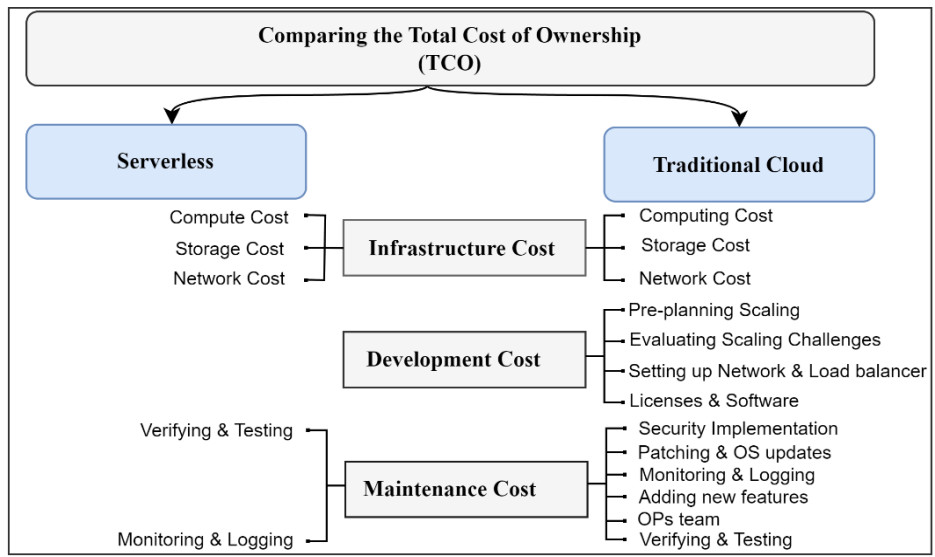

**Fig. 3.** A taxonomy of factors influencing the cost

*Development cost.* This refers to the effort and time spent designing and developing applications on cloud-based services. In traditional cloud, developers need to evaluate how the architecture would scale over time. The developer must focus on utilizing the resources in scaling up and down in a traditional cloud environment. Developers utilizing EC2 instances are required to dedicate significant time to assess potential scalability challenges within the IT architecture and decide on necessary tradeoffs in the preliminary stages. This incurred the cost of planning the resources and time. In addition, the developer must spend more time setting up a network, load balancer, purchasing licenses and software, and planning availability. In contrast, serverless computing leads developers to build the application without worrying about planning scaling and the deployment of the application. The cost of planning has become negligible in serverless computing.

*Maintenance cost.* This pertains to the ongoing cost required for running and maintaining an application. In the serverless, developers or operation teams do not need to maintain the application (e.g., patching and operating system updates). However, applications developed using cloud containers require extra work and labor to handle the application (i.e., DevOps team). The maintenance and operational costs become negligible in serverless computing compared to traditional cloud servers. Thus, leading to significantly lower costs overall and reducing the scalability issues and operational overhead.

Organizations considering adopting serverless or traditional cloud need to evaluate each component to make informed decisions.

## 6      Threat to Validity

Several potential threats could impact the validity of the results of this study. These threats are typically categorized into four primary categories: internal validity, construct validity, external validity, and conclusion validity [21].

**Internal validity:** refers to the degree to which specific factors influence methodological robustness. The first threat to this study is the participants' understanding of the interview questions. To mitigate this threat, we conducted pilot interviews with professionals from our network and provided them interview questions in advance. This ensured that the questions were both understandable and readable. We revised the interview questions based on the participants' feedback. The final interview preamble is provided in this study.

**Construct validity:** refers to the degree to which the research constructs are adequately substantiated and interpreted. The core constructs are the interview participants' viewpoints on the migration or adoption of serverless technology in the context of cost. The verifiability of the construct is considered the limitation of thematic analysis. Therefore, we followed a rigorous and step-by-step research method process and gave examples in quotations from the collected data (e.g., interviews).

**External validity** refers to the generalizability of the results. The sample size and sampling approach of this study may not generalize the findings. A common threat can arise that serverless is not widely adopted in the industry. Similarly, migration to serverless is not well established in the practice. Finding the potential sample size was challenging for us. We mitigated this threat by using possible sources such as social media platforms (e.g., LinkedIn, ResearchGate) and attending seven industrial meetups to find the potential population. We collected data from 4 countries across two continents from participants with diverse experience in various industrial domains and in serverless.

**Conclusion Validity:** refers to the factors that impact the trustworthiness of the study conclusion. To mitigate this threat, we conducted weekly meetings to develop the interview instruments and data analysis process. We reviewed the data based on the weekly discussion to improve the analysis process. Finally, we conducted a brainstorming session to draw the findings and conclusion of this study.

## 7      Conclusion and Future Work

Serverless computing presents a promising avenue for organizations to optimize costs and improve efficiency by minimizing scalability issues and operational overhead. However, successfully transitioning to serverless computing requires a deep understanding of cost implications and workload suitability. To this end, our study comprehensively analyzes cost optimization and workload suitability in serverless computing. Through an empirical investigation of eight systems and 15 interviews with industry professionals, we identified how companies predict the cost of adopting serverless, workload suitability, and factors that affect the cost of serverless applications. Furthermore, we presented a theoretical model for understanding the cost of serverless compared with traditional cloud.

Our study revealed that most of the organizations do unit cost economics and migrating legacy components to serverless to understand the cost benefits of serverless. Moreover, most of the systems and interviewers stated that serverless is suitable for highly predictable workload, where developers need to spend most of the time provisioning the underlying infrastructure. Three interviews stated that, while serverless theoretically offers infinite scalability, it has a burst concurrency limit that could not be cost-effective for certain stabilized high-scale applications. However, all the suggested developing greenfield projects with the serverless first mindset. Further they assisted transitioning to containers when it becomes more cost-effective. In addition, this study also identified factors that can increase the cost and strategies used to optimize the application cost. Finally, we developed a taxonomy for evaluating the cost of serverless versus traditional cloud computing. This taxonomy serves as a valuable tool for organizations, helping them make more informed decisions about which cloud computing model is most cost-effective for their specific needs.

As future work, we plan to extend our findings by mining Q&A repositories and conducting a survey with a larger number of industrial practitioners. Further, we aim to develop a comprehensive theory that explains how decisions are made at every stage of migrating to serverless computing—from planning and development to deployment.